\documentclass[aps,pra,twocolumn]{revtex4-1}

\usepackage[normalem]{ulem}
\usepackage{amsmath,amssymb}
\usepackage{multirow}
\usepackage{xcolor,soul}
\usepackage{srcltx}
\usepackage{hyperref,graphicx}

\newcommand{\Polimi}{$^1$Dipartimento di Fisica - Politecnico di Milano, Piazza Leonardo da Vinci, 32, I-20133 Milano, Italy}

\newcommand{\IFNCNR}{$^2$Istituto di Fotonica e Nanotecnologie - Consiglio Nazionale delle Ricerche, Piazza Leonardo da Vinci, 32, I-20133 Milano, Italy}

\newcommand{\RiceECE}{$^3$Department of Electrical and Computer Engineering, Rice University, 6100 Main Street, Houston, TX 77005}

\newcommand{\Rice}{$^4$Department of Physics and Astronomy, Laboratory for Nanophotonics, Rice University, 6100 Main Street, Houston, TX 77005}

\newcommand{\IIT}{$^5$Istituto Italiano di Tecnologia, via Morego 30, I-16163, Genova, Italy}

\newcommand{\China}{$^6$Cixi Institute of Biomedical Engineering, Ningbo Institute of Industrial Technology, Chinese Academy of Sciences, 1219 Zhongguan West Road, Ningbo 315201, China}

\begin{document}

\title{Transient optical symmetry breaking for ultrafast broadband dichroism\\ in plasmonic metasurfaces}

\author{Andrea Schirato$^{1,5^\star}$}

\author{Margherita Maiuri$^{1,2^\star}$}

\author{Andrea Toma$^5$}

\author{Silvio Fugattini$^5$}

\author{Remo Proietti Zaccaria$^{5,6}$}

\author{Paolo Laporta$^{1,2}$}

\author{Peter Nordlander$^{3,4}$}

\author{Giulio Cerullo$^{1,2}$}

\author{Alessandro Alabastri$^{3^\otimes}$}

\author{Giuseppe Della Valle$^{1,2^\otimes}$}

\address{\Polimi,\IFNCNR,\RiceECE,\Rice,\IIT,\China}

\maketitle

{\bf 
Ultrafast nanophotonics is an emerging research field aimed at the development of nanodevices capable of light modulation with unprecedented speed. A promising approach exploits the optical nonlinearity of nanostructured materials (either metallic or dielectric) to modulate their effective permittivity via interaction with intense ultrashort laser pulses. While the ultrafast temporal dynamics of such nanostructures following photoexcitation has been studied in depth, sub-ps transient spatial inhomogeneities taking place at the nanoscale have been so far almost ignored. Here we theoretically predict and experimentally demonstrate that the inhomogeneous space-time distribution of photogenerated hot carriers induces a transient symmetry breaking in a plasmonic metasurface made of highly symmetric metaatoms. The process is fully reversible, and results in a broadband transient dichroic optical response with a recovery of the initial isotropic state in less than 1 picosecond, overcoming the speed bottleneck caused by slower relaxation processes, such as electron-phonon and phonon-phonon scattering. Our results pave the way to the development of ultrafast dichroic devices, capable of Tera bit/s modulation of light polarization.
}\par

In the last decade, engineered optical nanomaterials, especially in two-dimensional configurations, known as metasurfaces, have emerged as a new platform for the manipulation and steering of light beams~\cite{Kildishev_Sci_2013, Yu_NatMat_2014, Kuznetsov_Sci_2016, Fan_LSA_2019, Neshev_LSA_2018}. In particular, recent studies have demonstrated the capability to exploit {\it nonlinear} nanostructures and metasurfaces illuminated by intense femtosecond laser pulses for the {\it ultrafast} control of light (see, e.g., Refs.~\citenum{Vasa_LPR_2009, Neshev_LSA_2018, Makarov_LPR_2017, Ren_Adv_Mat_2019} and references therein). A variety of configurations has been reported and the temporal dynamics following photoexcitation has been the subject of intense experimental and theoretical research. This is especially true for plasmonic nanostructures, where the nonlinear mechanism presiding over the all-optical control is a delayed third-order process, whose dynamics is dictated by energy transfer between internal degrees of freedom of the metallic nanosystem. Upon photoexcitation and following plasmon dephasing, three main steps take place on well separated time scales~\cite{Nordlander_NatNano_2015, Liu_ACSP_2018, Ren_Adv_Mat_2019, Wang_JPCC_2015, Baida_PRL_2011,  Della_Valle_PRB_2012, Sun_PRB_1994}: (i) electron equilibration, following electron-electron scattering events within few hundreds femtoseconds; (ii) electron-lattice equilibration, occurring in few picoseconds; (iii) the much slower (nanosecond) heat release to the environment via phonon-phonon scattering. The latter two processes, which affect both metallic and dielectric structures, pose a substantial limit to the development of realistic all-optical nanodevices capable of ultrafast modulation speed. The attempts reported so far to overcome this limitation managed to suppress only one of the two contributions from the slower dynamical processes, and have only been demonstrated on narrow wavelength bands~\cite{Shcherbakov_NL_2015, Della_Valle_ACSP_2017}. The feasibility of broadband all-optical modulation of light with full return to zero in less than 1 picosecond thus remains the open challenge of ultrafast nanophotonics. Also, among the different kinds of all-optical functionalities, ultrafast polarization switching has recently attracted a huge interest for advanced applications in photonics and beyond~\cite{Yang_NatPhot_2017, Nicholls_NatPhot_2017}. 

To tackle this challenge, we exploit a phenomenon so far almost neglected, that is the onset of ultrafast nanoscale spatial inhomogeneities at very early times following photoexcitation~\cite{Rudenko_AOM_2018}. We demonstrate that such spatio-temporal transients can break the symmetry of a plasmonic metasurface even if it consists of highly symmetric metaatoms, thus inducing a broadband dichroic response with ultrafast recovery of the initial isotropic configuration well before the complete relaxation of the system.

\begin{figure*}[htp!]
\includegraphics[width=16cm]{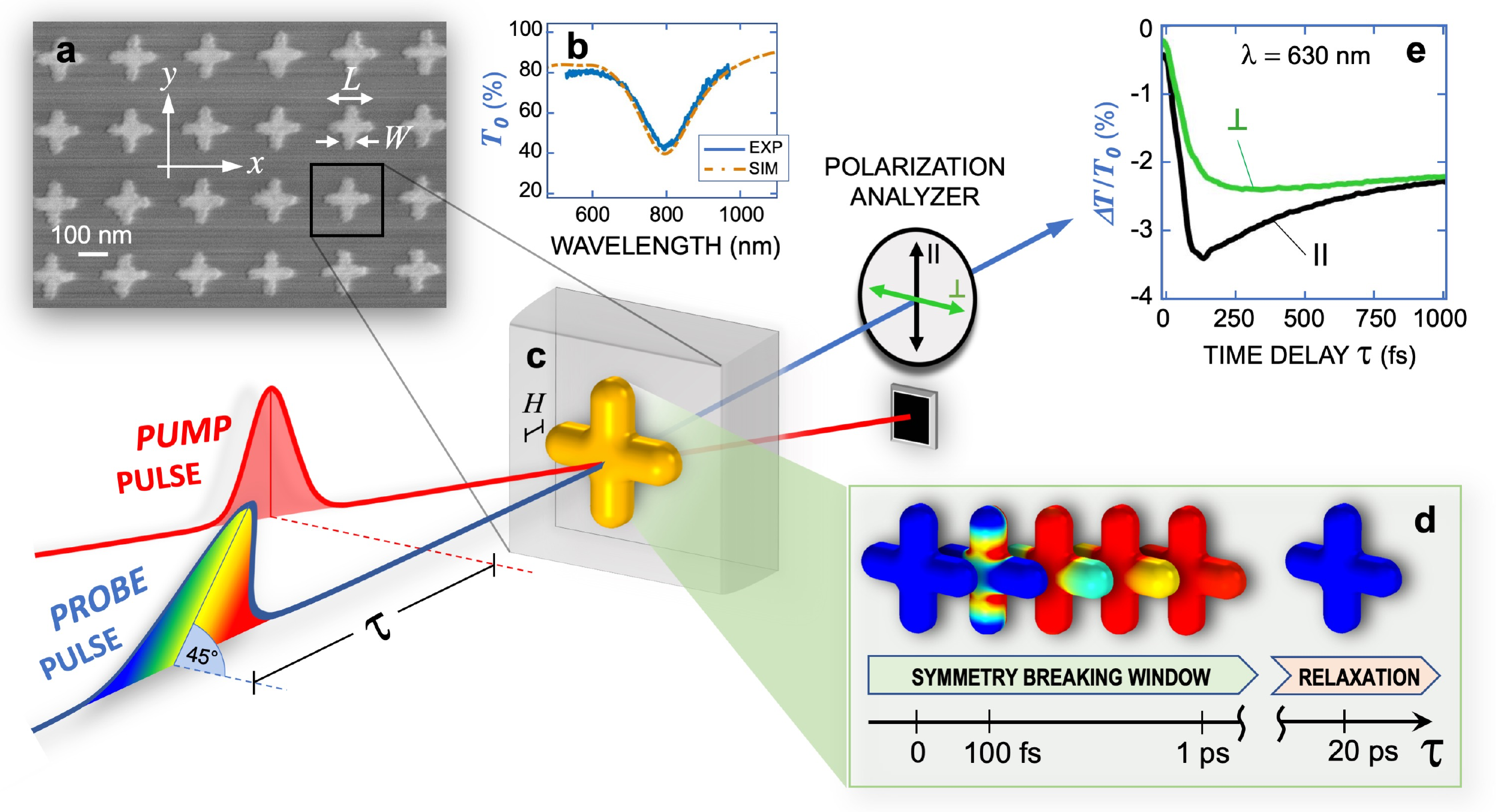}
\caption{{\bf Ultrafast optical dichroism in Au metasurface}. {\bf a}, SEM image of the metasurface. {\bf b}, Measured (blue) and simulated (orange) transmittance of the polarization insensitive unperturbed structure. {\bf c}, Sketch of the polarization-resolved pump-probe set-up utilized to experimentally reveal the ultrafast optical dichroism. {\bf d}, Cartoon of the transient permittivity pattern ($|\Delta\varepsilon''(\vec{r},\tau)|$, evaluated at around 630 nm) evolving over time at the nanoscale. {\bf e}, Experimental differential transmission signal at 630 nm after photoexcitation when the probe polarization is parallel (black) or perpendicular (green) to the pump field.
}
\end{figure*}

Our concept is illustrated in Figure~1. We consider a plasmonic metasurface made of a square array of closely packed ($\sim270$~nm periodicity) C4 symmetric gold nanocrosses with thickness $H=45$~nm, width $W=60$~nm and length $L=165$~nm (Fig.~1a). Such a symmetric nanomaterial provides a polarization independent static transmittance at normal incidence, $T_{0}$, characterized by a broad dip around 800 nm (Fig.~1b, solid curve). This dip is due to the degenerate longitudinal plasmonic resonances of the two arms of the nanocross and is broadened by hybridization effects in the array configuration. Such a degeneracy can be broken by the resonant absorption of an ultrashort pump pulse with linear polarization parallel to the direction of one of the arms (Fig.~1c). Photoexcitation creates a highly inhomogeneous near field, mostly because of the retardation-based nature of plasmonic resonances in relatively large nanostructures. The inhomogeneous absorption pattern in each metaatom locally affects the electronic energy distribution of gold, inducing a non-uniform out of equilibrium hot-carriers distribution which anisotropically modifies the metal permittivity (Fig.~1d). The fingerprint of the ultrafast pump-induced symmetry breaking is a transient transmission anisotropy (Fig.~1e), which can be revealed in a polarization-resolved pump-probe experiment where the delayed probe pulse impinges at normal incidence with a linear polarization at 45$^\circ$ to the nanocross arms (Fig.~1c).

We first developed a quantitative numerical model able to describe the ultrafast spatial transients in terms of three space- and time-dependent variables detailing the internal energy dynamics in gold: the energy density stored in the nonthermal fraction of photoexcited carriers, $N(\vec{r},t)$, the temperature of the thermalized population of hot electrons, $\Theta_e(\vec{r},t)$, and the lattice temperature, $\Theta_l(\vec{r},t)$ (which plays a minor role on the considered sub-picosecond time scales). These three variables are interlinked by a set of three coupled partial differential equations (detailed in the Methods section), which we refer to as the Inhomogeneous Three-Temperature Model (I3TM). The I3TM extends the popular 3TM~\cite{Sun_PRB_1994}, widely employed so far both for metallic and semiconducting nanostructures (see, e.g., Refs.~\citenum{Zavelani_ACSP_2015, Gaspari_NL_2017}) by including the spatial dependence of the variables. The source term is the instantaneous absorbed power density $P_{abs}(\vec{r},t)$, driving the photo-excited nonthermal carriers, given by the equation
\begin{equation}\label{eq_drive}
P_{abs}(\vec{r},t) = \frac{F S}{V} A(\vec{r}) g(t),
\end{equation}
where $g(t)$ is the normalized Gaussian intensity profile of the ultrashort pump pulse, $A(\vec{r})$ the inhomogeneous absorption pattern of the pump on the individual metaatoms (see Methods for details), $F$ the pump pulse fluence, $V$ the volume of the metaatom, and $S$ the area of the unit cell of the metasurface.

\begin{figure*}[ht!]
\includegraphics[width=15cm]{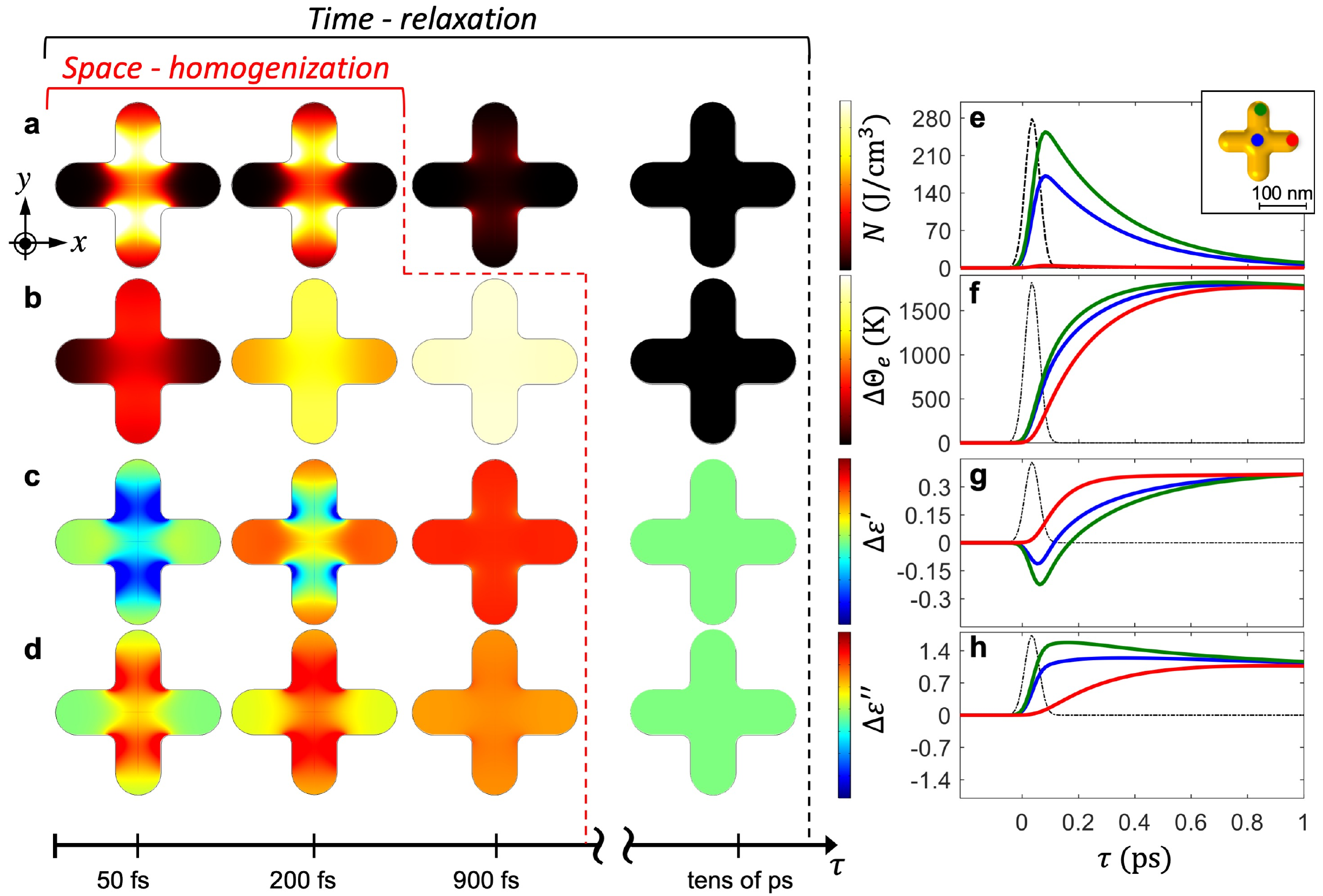}
\caption{{\bf Theoretical modelling}. Theoretical results for space-time dependent ultrafast nonlinearities in plasmonic nanoparticles. {\bf a - d}, From top to bottom: space-time dynamics of  the energy density stored in the nonthermal fraction of hot electrons population, $N$ (J/cm$^3$); increase of electronic temperature, $\Delta \Theta_e$ (K); variation of real part, $\Delta \varepsilon'$, and imaginary part, $\Delta \varepsilon''$, of dielectric permittivity in the plasmonic nanocross (here shown at 630 nm), for different pump-probe delays (pump fluence $F = 300$~$\mu$J/cm$^2$). The maps are taken in the central cross-section of the irradiated nanoparticle (i.e. the $z = H/2$ plane). {\bf e - h}, From top to bottom: corresponding time evolution of the variables mapped in (a)-(d) evaluated at three characteristic positions visualized in the inset of panel (e). The dash-dotted curve represents the temporal profile $g(t)$ of the pump pulse (see Methods).}
\end{figure*}

Figure~2a shows the $x,y$ cross-section of the energy density $N$ of nonthermal carriers, numerically calculated at selected time delays after excitation with a pump pulse at 860 nm wavelength (i.e.~slightly red shifted with respect to the plasmonic resonance peak), linearly polarized along the vertical $y$-axis. Note that the pattern distribution is highly asymmetric, with a major localization along the vertical arm of the nanocross which corresponds to the polarization of the excitation beam. The ultrafast decay of $N$ (Fig.~2e) follows the electron-electron scattering dynamics, with a characteristic time constant of about $350$~fs, as estimated from more accurate theoretical models~\cite{Zavelani_ACSP_2015, Sun_PRB_1994}. The subsequent energy release to the thermalized hot carriers gives rise to an inhomogeneous pattern also for the electronic temperature $\Theta_e$ (Fig.~2b). 

The thermal hot electron distribution eventually undergoes a spatial diffusion process governed by the thermal conductivity of the Fermi gas $\kappa(\Theta_e)\propto \Theta_e$~\cite{Kanavin_PRB_1998}, which restores a homogeneous distribution of hot electrons in few hundreds femtoseconds (Fig.~2f), thus closing the temporal window for the photoinduced electronic symmetry breaking. Note that electron-lattice equilibration, leading to the complete relaxation of the electronic subsystem, takes place over a much longer timescale of a few picoseconds, and thus does not play any role in the transient anisotropy, similarly to the even longer dynamics of heat release from the metal lattice to the environment. 

Starting from the simulated $N(\vec{r},t)$ and $\Theta_e(\vec{r},t)$ distributions, we calculated the subsequent modulation of gold permittivity, $\Delta\varepsilon(\vec{r},t)$, following the same approach described in previous works~\cite{Zavelani_ACSP_2015}, but now taking into account the spatial dependence (see Methods). Note that $\Delta\varepsilon(\vec{r},t)$ inherits a highly asymmetric distribution on the sub-picosecond time scale, mostly reminescent of the pattern of $N$, even though the temporal dynamics of Figs.~2g-2h exhibit more complex features, due to the interplay between the contributions arising from the two different (nonthermal and thermal) electronic populations. This makes the transient symmetry breaking of the spatial electronic distribution internal to the nanosystem readily accessible by optical means in terms of an ultrafast polarization sensitive optical response. 

We employed finite-element method numerical analysis in the frequency domain to calculate the transmission spectrum $T_{||,\bot}(\lambda,\tau)$ of the metasurface at normal incidence for a linear polarization either parallel ($||$) or orthogonal ($\bot$) to the pump, for a set of permittivity configurations obtained by sampling the $\Delta\epsilon(\vec{r},t=\tau)$ distribution at different values of the pump-probe delay $\tau$. The relative differential transmittance for each polarization, normalized to the static transmittance of the unperturbed structure, is then determined as 
\begin{equation}
\Delta_{||,\bot} = \frac{\Delta T_{||,\bot} }{T_{0}} = \frac{T_{||,\bot}(\lambda,\tau)}{T_{0}(\lambda)}-1.
\end{equation}
An ultrafast broadband dichroism should manifest itself as a difference between $\Delta_{\|}$ and $\Delta_{\bot}$, for a wide range of wavelengths, and then disappear within less than 1 ps.

\begin{figure*}
\includegraphics[width=16.5cm]{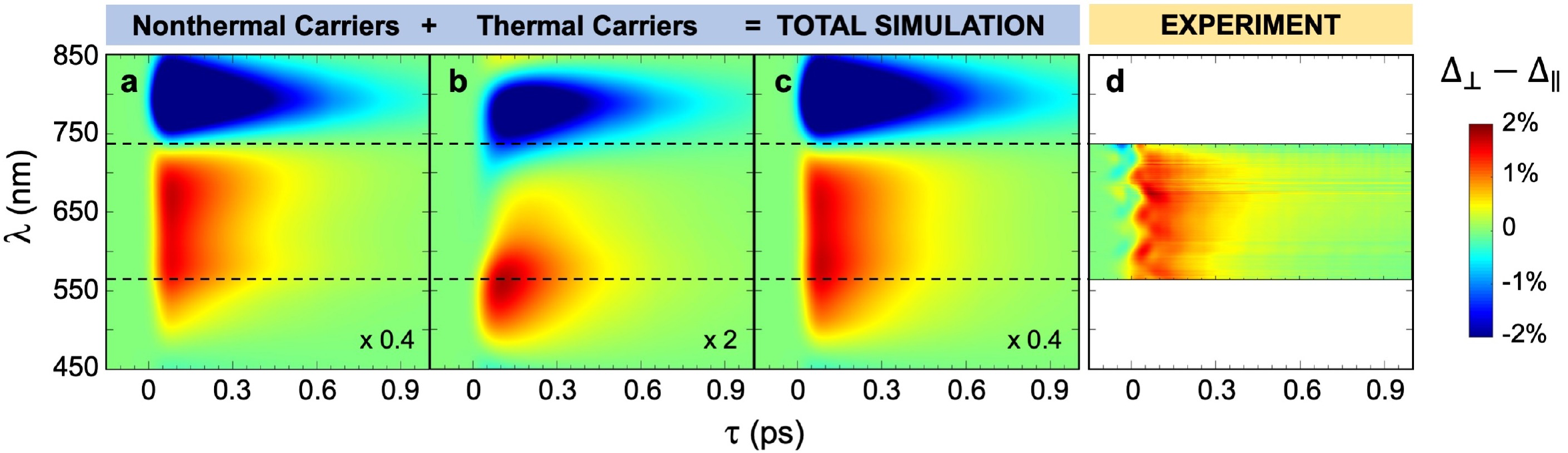}
\caption{{\bf Broadband ultrafast dichroic transmittance: theory and experiments.} {\bf a - c}, Theoretical prediction of the ultrafast photo-induced dichroic spectrum, $\Delta_{\bot}-\Delta_{\|}$ (according to Eq.~2), at different pump-probe delays, disentangled in terms of the contributions arising from nonthermal carriers (a), and thermalized carriers (b), together with the total effect (c). {\bf d}, Corresponding pump-probe measurement in the  spectral range accessible by the experiment (fluence $F \simeq 400$~$\mu$J/cm$^2$).}
\end{figure*}

\begin{figure}[hb!]
\includegraphics[width=8.5cm]{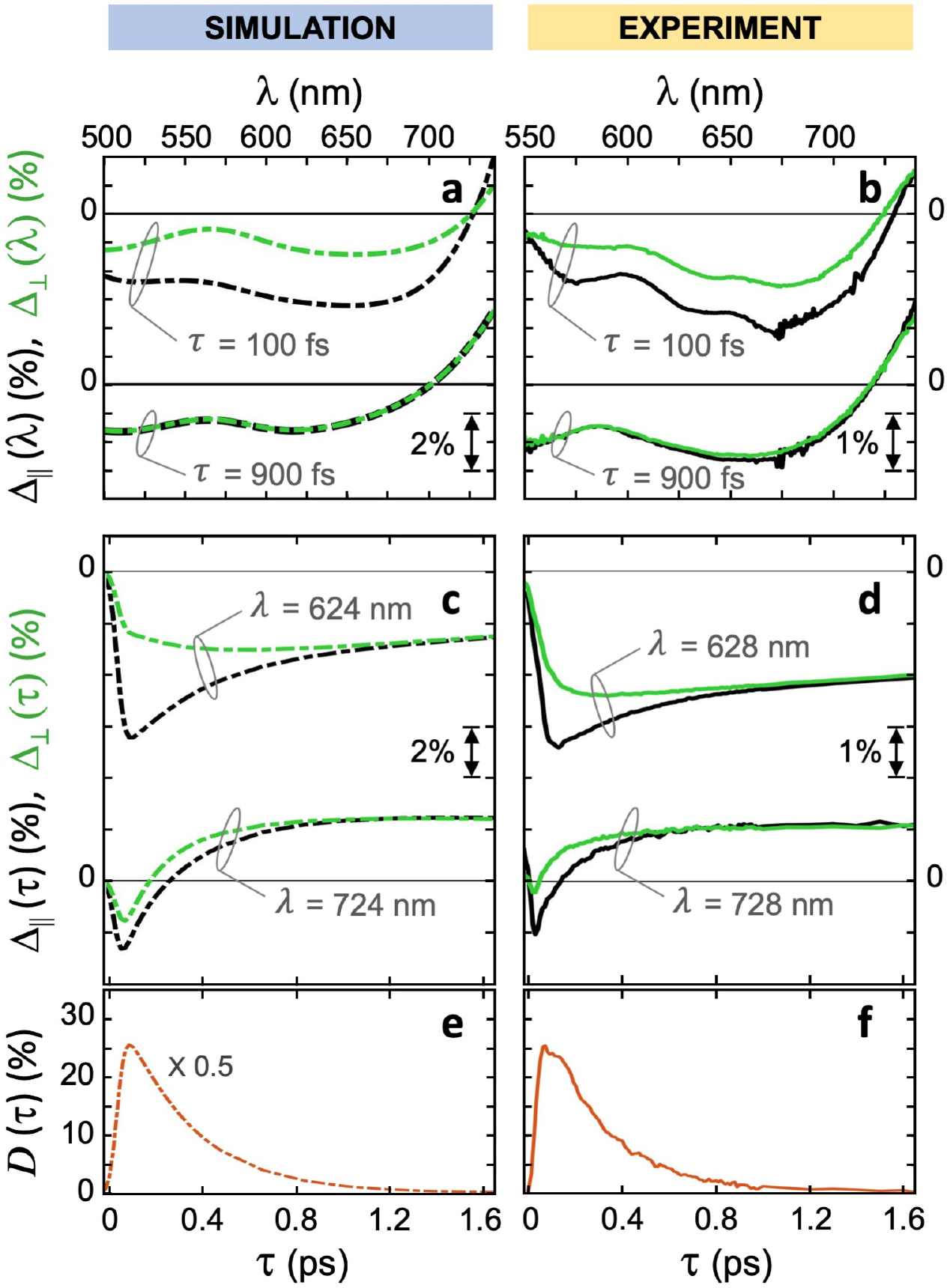}
\caption{{\bf Symmetry Breaking Window.} Comparison between theoretical predictions (left, dash-dotted lines) and experimental data (right, solid lines) of the optical symmetry breaking in Au nanocrosses. {\bf a - b}, Polarization-resolved relative differential transmission spectra, $\Delta_{\|,\bot}$, at two different time delays: probe polarisation parallel (black) and orthogonal (green) to the pump. {\bf c - d}, Time evolution of the $\Delta_{\|,\bot}$, at two fixed wavelengths. {\bf e - f}, Time evolution of the ultrafast dichroic ratio, $D$, at around 620 nm.}
\end{figure}

Figures 3a-3c show the results of the simulations, where $\Delta_{\bot}-\Delta_{\|}$ is calculated from 450 nm to 850 nm. Note that the main contribution to the phenomenon arises from nonthermal carriers (Fig.~3a). However, thermalized carriers (Fig.~3b) also contribute but to a lesser degree, which is remarkable given that the electronic temperature relaxation takes place on a much longer time scale, far beyond the symmetry breaking window (Fig.~2b and 2f).

Our theoretical predictions are compared with polarization resolved ultrafast pump-probe experiments, performed according to the sketch of Fig.~1c. The experimental transient linear dichroism $\Delta_{\bot}-\Delta_{\|}$ is shown in Fig.~3d and well matches the simulations of Fig.~3c on the relevant spectral range (565-735 nm), apart from a scaling factor of about 2, which is consistent with the fact that semi-classical calculations (as in our I3TM) tend to overestimate the generation of high energy nonthermal electrons compared to more rigorous quantum models \cite{Manjavacas_ACSNANO_2014, Besteiro_ACSP_2017, Besteiro_NanoToday_2019}. Moreover, the sign switch at longer wavelengths (750-850 nm) could not be probed experimentally, but is consistent with  transient absorption spectra of gold nanostructures \cite{Sun_PRB_1994, Baida_PRL_2011, Zavelani_ACSP_2015, Wang_JPCC_2015}. The complex spectral oscillations of the map at the very early stages of the dynamics ($-60 \lesssim \tau \lesssim 60$~fs) are due to instrumental artifacts caused by unavoidable inter-pulse four-wave mixing in the dielectric substrate (see, e.g. Ref.~\citenum{Dietzek_LPL_2007} and references therein) but do not prevent the observation of the transient symmetry-breaking.
A more direct comparison between theory and experiments is provided in Fig.~4, where we show the simulated (Fig.~4a) and the measured (Fig.~4b) spectra of $\Delta_{\|}$ (black traces) and $\Delta_{\bot}$ (green traces) at two different time delays. A good agreement is retrieved also for the dynamics of the $\Delta_{\|,\bot}$ (compare Fig.~4c and 4d), including the ultrafast sign change in the red wing of the spectrum, i.e.~at around $720-730$~nm (compare lower traces in Fig.~4c and 4d). Finally, since the dichroic contrast $\Delta_{\bot}-\Delta_{\|}$ of Fig.~3 scales with the pump fluence, we introduced a transient dichroic ratio, $D = (\Delta_{\bot}-\Delta_{\|})/\max\{|\Delta_{\bot}+\Delta_{\|}|\}$, which   quantifies the ultrafast dichroic performance {\it intrinsic} to the considered structure. Note that this figure of merit reaches values as high as few tens \% at the peak of the dichroic dynamics, at around 100 fs time delay (Figs.~4c-4d).

To summarize, we have shown that the absorption of intense ultrashort pulses induces a transient symmetry breaking in a plasmonic metasurface with C4 symmetric metaatoms. Such spatio-temporal electronic transients at the nanoscale translate into an anisotropic local permittivity distribution with C2 symmetry, and subsequent ultrafast dichroic optical response of the metasurface, dominated by nonthermal carriers dynamics. This modulation is broadband and returns to the isotropic configuration within few hundreds femtoseconds. We envisage that the transient spatial inhomogeneities of hot electrons revealed in our experiments can disclose an unprecedented route for the all-optical control of light at Tera bit/s speed, with particular relevance for ultrafast polarization management~\cite{Yang_NatPhot_2017, Nicholls_NatPhot_2017}. Our results can also pave the way to the development of an optimized platform for hot-electrons-based photocatalysis~\cite{Linic_NatMat_2015}.

\section*{Data availability}
The data that support the plots within this paper and other findings of this study are available from the corresponding authors upon reasonable request.
{\small 
\acknowledgements{We acknowledge financial support from Graphene FET Flagship Core Project 3, Grant No. 881603. G.D.V and G.C. acknowledge the Project METAFAST-899673-FETOPEN-H2020. G.D.V. and A.S. acknowledge support from the Italian MIUR under the PRIN Grant No.~2015WTW7J3. P.N. acknowledges support from the Robert A. Welch Foundation under grant C-1222.}

\section*{Author contributions}
\noindent 
G.D.V. and G.C. conceived and designed the experiment. A.T., S.F. and R.P.Z. manufactured the samples and performed the static measurements. A.S., A.A., P.N. and G.D.V. developed the theory and designed the structures. A.S. performed the numerical simulations.  M.M. performed the pump-probe experiment. G.C., R.P.Z. and P.L. supervised the experimental work. G.D.V., A.A. and A.S. wrote the first draft of the manuscript. All authors discussed the results and commented on the manuscript.\\

\noindent $^\star$A.S. and M.M. equally contributed to this work. \\
$^\otimes$A.A. and G.D.V. equally contributed to this work. \\
\href{mailto:alessandro.alabastri@rice.edu}{alessandro.alabastri@rice.edu}, \href{mailto:giuseppe.dellavalle@polimi.it}{giuseppe.dellavalle@polimi.it} 

\section*{Additional information}
Supplementary information is available in the online version of the paper.

\section*{Competing financial interests}
The authors declare no competing financial interests.

\section*{Methods}

{\bf Theory and numerical modelling.} The I3TM reads as follows:
\begin{eqnarray}
\frac{\partial N }{\partial t} &=&  -a N - bN + P_{abs}(\vec{r},t), \label{eq:inhom3tm_1}\\
\frac{\partial}{\partial t} C_e \Theta_e &=& - \nabla \Big(- \kappa_e \nabla \Theta_e \Big) - G \Big(\Theta_e - \Theta_l \Big) +a N, \label{eq:inhom3tm_2}\\
C_l \frac{\partial \Theta_l}{\partial t} &=& \kappa_l \nabla^2 \Theta_l + G \Big(\Theta_e - \Theta_l \Big) + bN, \label{eq:inhom3tm_3}
\end{eqnarray}
\noindent
where the explicit dependence on both time and space for $N(\vec{r},t)$, $\Theta_e(\vec{r},t)$ and $\Theta_l(\vec{r},t)$ have been dropped for the sake of conciseness. 
$a$ and $b$ represent the out-of-equilibrium electron gas heating rate and the nonthermal electron-phonon scattering rate, respectively. They are related to the electron-electron ($\tau_{e-e}$) and the electron-phonon ($\tau_{e-ph}$) scattering times: $a=\langle 1/\tau_{e-e} \rangle = (\hbar \omega_P)^2 / 2 \tau_0 E_F^2$, with $\hbar \omega_P$ the pump photon energy, $\tau_0=13.5$ fs and $E_F$ the Fermi energy in Au; $b=1/\tau_{e-ph} = k_B \Theta_D / (\tau_f \hbar \omega_P)$, where $k_B$ is the Boltzmann constant, $\Theta_D$ the metal Debye temperature, and $\tau_f$ the quasi-particle free flight time (see Ref.~\citenum{Zavelani_ACSP_2015} and references therein). Note that for $\tau_0$ we doubled the value compared to the standard 3TM, as it is well known that the latter underestimates the effective $\tau_{e-e}$, compared to more refined models (see, e.g., Ref.~\citenum{Della_Valle_PRB_2012}). The other coefficients are the electron (lattice) specific heat, $C_e$ ($C_l$), and thermal conductivity, $\kappa_e$ ($\kappa_l$), and the electron-photon coupling constant, $G$. Concerning $C_e=C_e(\Theta_e)$ and $G = G(\Theta_e)$, these are taken from Ref.~\cite{Lin_PRB_2008}, where Density Functional Theory (DFT) is applied to determine the total density of electronic states including both conduction and valence bands. Electron diffusion is instead ruled by a conductivity $\kappa_e = \kappa_e(\Theta_e) = \kappa_l \Theta_e/\Theta_l$, as in Ref.~\citenum{Kanavin_PRB_1998}, with $\kappa_l=316$ W m$^{-1}$ K$^{-1}$ \cite{Ashcroft_1960}. Finally, $C_L=2.49 \cdot 10^6$ J m$^{-3}$ K$^{-1}$, after Ref.~\citenum{Lide_1993}.

Regarding the source term, its temporal evolution is written as $g(t) = \sqrt{4 \ln 2 / (\pi \Delta t^2)} \exp[-4 \ln 2 (t-t_0)^2/\Delta t^2]$, with $\Delta t$ the pulse duration (full width at half maximum intensity). We assumed $\Delta t = 60$~fs in order to take into account the typical stretching of the signal dynamics induced by the coherent artifact~\cite{Dietzek_LPL_2007}. The absorption pattern $A(\vec{r})$ is calculated at the peak wavelength of the pump spectrum ($\lambda_P = 860$~nm) by 3D finite element method (FEM) numerical analysis, employing a commercial tool (COMSOL Multiphysics 5.4). The geometrical parameters used in the simulations ($L = 155$~nm, $W = 40$~nm, $H = 30$~nm, and $285$~nm array period) were adapted from the nominal values (estimated from SEM analysis) in order to best fit the experimental transmission spectrum of the unperturbed structure (compare solid curve and dash-dotted curve in Fig.~1b). We also assumed smoothed edges of the nanocross with 20/8~nm radius of curvature for outer/inner corners.
Port formalism in the frequency domain with periodic boundary conditions was adopted, so to simulate an infinite square array of nanocrosses, with plane wave excitation at normal incidence, linearly polarized along one of the nanocross arms ($y$-axis in Fig.~1c). The geometrical domains were finely meshed in order to resolve the effective (local) optical wavelength with at least 7 elements in the dielectric regions, whereas an even finer mesh was employed in gold (with linear element size comprised between $2.5$~nm and $10$~nm) so to correctly describe evanescent decay patterns in the nanocrosses.
The I3TM (Eqs.~\ref{eq:inhom3tm_1}-\ref{eq:inhom3tm_3}) is then locally solved in the time domain with the Generalised Alpha method, with a time step $dt=\Delta t/20$ (so to properly resolve the pump pulse driving dynamics). 

Once the spatio-temporal dynamics of the internal energy variables ($N$, $\Theta_e$ and $\Theta_l$) is computed, the corresponding inhomogeneous transient and dispersed permittivity perturbation $\Delta \varepsilon(\vec{r}, t, \lambda)=\Delta \varepsilon\Big( N(\vec{r},t),\Theta_e(\vec{r},t),\Theta_l(\vec{r},t),\lambda \Big) $ is then modeled according to the third-order delayed optical nonlinearity of noble metals detailed in our previous studies (see, e.g.~\cite{Della_Valle_PRB_2012, Zavelani_ACSP_2015} and references therein).  In short, the electronic excitation results into a variation of occupation probability for the energy levels in the conduction band. This in turns modulates the joint density of states for the dominant (valence-conduction) interband optical transition of gold, around the L point in the first Brillouin zone~\cite{Rosei_SS_1973}. The latter is modeled in the parabolic anisotropic approximation, with effective masses and energy constants taken from atomistic calculations~\cite{Christensen_PRB_1971}, and the corresponding variations of gold absorption coefficient, or imaginary part of material permittivity, is then retrieved in a broad band of wavelengths via semiclassical theory under constant matrix elements approximation (see Ref.~\citenum{Zavelani_ACSP_2015} and references therein). The variation of the real part of gold permittivity is finally computed by Kramers-Kronig analysis.

The final step, i.e. the modelling of the interaction between the photoexcited structure and a probe pulse at a given time delay, is then performed via a further frequency-domain analysis spanning the $450-850$~nm wavelength range for two orthogonal linear polarizations, being either parallel or orthogonal to the vertical axis of the nanocrosses (aligned with the pump polarization). Here we assumed a gold permittivity given by $\varepsilon_{Au}(\lambda,t) = \varepsilon_{Au}^{static}(\lambda)+\Delta \varepsilon(\lambda,t)$ with $t = \tau$ a parameter representing the pump-probe time delay of our experiments and $\varepsilon_{Au}^{static}(\lambda)$ the permittivity of gold at the equilibrium, provided by an analytical model~\cite{Etchegoin_JCP_2006} fitted on Johnson and Christy experimental data~\cite{Johnson_PRB_1972}.\\

{\bf Fabrication of the plasmonic metasurface.} A plasmonic metasurface consisting of closely packed nanocrosses was fabricated recurring to Electron Beam Lithography (EBL). The process was carried out on CaF$_2$ (100) substrates spin-coated at 1800 rpm with Poly(methyl methacrylate) (MicroChem 950 PMMA A2) electronic resist. An aluminum film of 10 nm thickness was thermally deposited onto the PMMA surface in order to avoid charging and drifting effects. Then, an EBL machine (Raith 150-two) equipped with a pattern generator, was operated for the nanostructure direct writing (electron energy 20 keV and beam current 28 pA). After the Al removal in a KOH-H$_2$O solution at a concentration of 1M, the exposed resist was developed in a conventional solution of methyl isobutyl ketone-isopropyl alcohol (MIBK-IPA) (1:3) for 30 s. To complete the development process and to prevent PMMA scum, the substrate was additionally immersed for 30 s in isopropyl alcohol. Physical vapour deposition (evaporation rate 0.3~${\rm \AA}$/s) respectively of 5 nm Ti as adhesion layer and 45 nm Au was performed on the sample. Finally, the unexposed resist was removed with acetone and rinsed out in IPA. O$_2$ plasma ashing at 100 W for 180 s was employed to remove residual resist and organic contaminants.

{\bf Pump-probe measurements.} The experimental setup used for high time resolution broadband pump-probe spectroscopy is described in detail elsewhere~\cite{Polli_RSI_2007}. An amplified femtosecond Ti:sapphire laser at 2 kHz repetition rate pumps two non-collinear optical parametric amplifiers (NOPAs), which generate the pump and probe pulses, respectively. The pump pulses, slightly red-shifted with respect to the longitudinal plasmonic resonance of the nanocross arms, cover the 850-950 nm bandwidth and are compressed by a fused silica prism pair to sub-30-fs duration; the probe pulses cover the 565-735 nm bandwidth and are compressed by chirped mirrors to sub-10-fs duration. Pump and probe propagate non-collinearly at an angle of $\sim5^\circ$ and are focused by a spherical mirror to a diameter of $50~\mu$m which is smaller than the sample area patterned with nanocrosses. The sample is aligned perpendicularly to the probe beam. The transmitted probe, spatially selected by an iris, is dispersed in a spectrometer and detected with a CCD with a custom electronics capable of reading out spectra at the full laser repetition rate. The pump pulse is modulated at 1 kHz by a mechanical chopper and the relative differential transmittance ($\Delta_{\|,\bot} = \Delta T_{\|,\bot} / T_{0}$) signal is measured as a function of probe wavelength $\lambda$ and pump-probe delay $\tau$. In order to avoid pump-probe misalignment caused by rotation of the probe polarization between the two orthogonal configurations, we kept the probe polarization fixed at an angle of $45^\circ$ with respect to the pump, which is in turns parallel to one of the arms of the nanocrosses, and performed polarization filtering at the output. This configuration is equivalent to the one considered in the numerical simulations in terms of the $\Delta_{\bot}$ and $\Delta_{\|}$, and guarantees a more reliable evaluation of the contrast between the two spectra.

\bibliography{MEP_BIB}

\begin{thebibliography}{}

\bibitem{Kildishev_Sci_2013}
Kildishev, A. V., Boltasseva, A., Shalaev, V. M. Planar Photonics with Metasurfaces. {\it Science} {\bf 339}, 1289 (2013).

\bibitem{Yu_NatMat_2014}
Yu, N., Capasso, F. Flat optics with designer metasurfaces. {\it Nat. Mater.} {\bf 13}, 139 (2014).

\bibitem{Kuznetsov_Sci_2016}
Kuznetsov, A. I., Miroshnichenko, A. E., Brongersma, M. L., Kivshar, Y. S., Luk’yanchuk, B. Optically resonant dielectric nanostructures. {\it Science} {\bf 354}, aag2472 (2016).

\bibitem{Fan_LSA_2019}
Phan, T., Sell, D., Wang, E. W., Doshay, S., Edee, K., Yang, J., Fan, J. A. High-efficiency, large-area, topology-optimized metasurfaces. {\it Light: Sci. and Appl.} {\bf 8}, 48 (2019).

\bibitem{Neshev_LSA_2018}
Neshev, D. N., Aharonovich, I. Optical metasurfaces: new generation building blocks for multi-functional optics. {\it Light: Sci. and Appl.}  {\bf 7}, 59 (2018).

\bibitem{Vasa_LPR_2009}
Vasa, P., Ropers, C., Pomraenke, R., Lienau, C. Ultra-fast nano-optics. {\it Laser \& Photon. Rev.} {\bf 3}, 483 (2009).

\bibitem{Makarov_LPR_2017}
Makarov, S. V., Zalogina, A. S., Tajik, M., Zuev, D. A., Rybin, M. V., Kuchmizhak, A. A., Juodkazis, S., Kivshar, Y. Light-Induced Tuning and Reconfiguration of Nanophotonic Structures. {\it Laser \& Photon. Rev.} {\bf 11}, 1700108 (2017).

\bibitem{Ren_Adv_Mat_2019}
Ren, M., Cai, W., Xu, J. Tailorable Dynamics in Nonlinear Optical Metasurfaces. {\it Adv. Materials}  1806317 (2019). DOI: 10.1002/adma.201806317

\bibitem{Nordlander_NatNano_2015}
Brongersma, M., Halas, N. J. \& Nordlander, P. Plasmon-induced hot carrier science and technology. {\it Nature Nanotech.} {\bf 10}, 25-34 (2015).

\bibitem{Liu_ACSP_2018}
Liu, J. G., Zhang, H., Link, S. \& Nordlander, P. Relaxation of Plasmon-Induced Hot Carriers. {\it ACS Photonics} {\bf 5} 2584-2595 (2018).

\bibitem{Wang_JPCC_2015}
Wang, X., Guillet, Y., Selvakannan, P. R., Remita, H., Palpant, B. Broadband Spectral Signature of the Ultrafast Transient Optical Response of Gold Nanorods. {\it J. Phys. Chem. C} {\bf 119}, 7416 (2015).

\bibitem{Baida_PRL_2011}
Baida, H., Mongin, D., Christofilos, D., Bachelier, G., Crut, A., Maioli, P., {Del Fatti}, N., Vall\'ee, F. Ultrafast Nonlinear Optical Response of a Single Gold Nanorod near Its Surface Plasmon Resonance. {\it Phys. Rev. Lett.} {\bf 107}, 057402 (2011).

\bibitem{Della_Valle_PRB_2012}
{Della Valle}, G., Conforti, M., Longhi, S., Cerullo, G., Brida, D. Real-time optical mapping of the dynamics of nonthermal electrons in thin gold films. {\it Phys. Rev. B} {\bf 15}, 155139 (2012).

\bibitem{Sun_PRB_1994}
Sun, C.-K., Vall\'ee, F., Acioli, L. H., Ippen, E. P., Fujimoto, J. G. Femtosecond-tunable measurement of electron thermalization in gold. {\it Phys. Rev. B} {\bf 50}, 15337 (1994).

\bibitem{Shcherbakov_NL_2015} 
Shcherbakov, M. R., Vabishchevich, P. P., Shorokhov, A. S., Chong, K. E., Choi, D.-Y., Staude, I., Miroshnichenko, A.E., Neshev, D. N., Fedyanin, A. A., Kivshar, Y. Ultrafast All-Optical Switching with Magnetic Resonances in Nonlinear Dielectric Nanostructures. {\it Nano Lett.} {\bf 15}, 6985-6990 (2015).

\bibitem{Della_Valle_ACSP_2017} 
Della Valle, G., Hopkins, B., Ganzer, L., Stoll, T., Rahmani, M., Longhi, S., Kivshar, Y. S., De Angelis, C., Neshev, D. N., and Cerullo, G. Nonlinear Anisotropic Dielectric Metasurfaces for Ultrafast Nanophotonics. {\it ACS Photonics} {\bf 4}, 2129-2136 (2017).

\bibitem{Yang_NatPhot_2017}
Yang, Y., Kelley, K., Sachet, E., Campione, S., Luk, T. S., Maria, J.-P., Sinclair, M.B., Brener, I. Femtosecond optical polarization switching using a cadmium oxide-based perfect absorber. {\it Nature Photonics} {\bf 11}, 390-396 (2017).

\bibitem{Nicholls_NatPhot_2017} 
Nicholls, L. H., Rodr\'{\i}guez-Fortu\~{n}o, F. J., Nasir, M. E., C\'ordova-Castro, R. M., Olivier, N., Wurtz, G. A. \& Zayats, A. V. Ultrafast synthesis and switching of light polarization in nonlinear anisotropic metamaterials. {\it Nature Photonics} {\bf 11}, 628-633 (2017).

\bibitem{Rudenko_AOM_2018} 
Rudenko, A., Ladutenko, K., Makarov, S., Itina, T. E. Photogenerated Free Carrier-Induced Symmetry Breaking in Spherical Silicon Nanoparticle. {\it Adv. Optical Mater.} {\bf 6}, 1701153 (2018).

\bibitem{Zavelani_ACSP_2015}
Zavelani-Rossi, M., Polli, D., Kochtcheev, S. Baudrion, A.-L., B\'{e}al, J., Kumar, V., Molotokaite, E., Marangoni, M., Longhi, S., Cerullo, G., Adam, P.-M., Della Valle, G. Transient optical response of a single gold nanoantenna: The role of plasmon detuning. {\it ACS Photonics} {\bf 2} 521-529 (2015).

\bibitem{Gaspari_NL_2017}
Gaspari, R., Della Valle, G., Ghosh, S., Kriegel, I., Scotognella, F., Cavalli, A., Manna, L. Quasi-Static Resonances in the Visible Spectrum from All-Dielectric Intermediate Band Semiconductor Nanocrystals {\it Nano Lett.} {\bf 17}, 7691 (2017).

\bibitem{Kanavin_PRB_1998}
Kanavin, A. P., Smetanin, I. V., Isakov, V. A., Afanasiev, Y. V., Chichkov, B. N., Wellegehausen, B., Nolte, S., Momma, C., T\"unnermann, A. Heat transport in metals irradiated by ultrashort laser pulses. {\it Phys. Rev. B} {\bf 57}, 698-703 (1998).

\bibitem{Dietzek_LPL_2007}
Dietzek, B, Pascher, T., Sundstr\"{o}m, T and Yartsev, A. Appearance of coherent artifact signals in femtosecond transient absorption spectroscopy in dependence on detector design. {\it Laser Phys. Lett.} {\bf 4}, 38 (2007).

\bibitem{Manjavacas_ACSNANO_2014}
Manjavacas, A., Liu, J. G., Kulkarni, V., Nordlander, P. Plasmon-Induced Hot Carriers in Metallic Nanoparticles. {\it ACS Nano} {\bf 8}, 7630 (2014).

\bibitem{Besteiro_ACSP_2017} 
Besteiro, L. V., Kong, X.-T., Wang, Z., Hartland, G., Govorov, A. O. Understanding Hot-Electron Generation and Plasmon Relaxation in Metal Nanocrystals: Quantum and Classical Mechanisms. {\it ACS Photon.} {\bf 4}, 2759 (2017).

\bibitem{Besteiro_NanoToday_2019}
Besteiro, L. V., Yua, P., Wang, Z., Holleitner, A. W., Hartland, G. W., Wiederrecht, G. P., Govorov, A. O. The fast and the furious: Ultrafast hot electrons in plasmonic metastructures. Size and structure matter. {\it Nano Today} {\bf 27}, 120 (2019).

\bibitem{Linic_NatMat_2015}
Linic, S., Aslam, U., Boerigter, C., Morabito, M. Photochemical transformations on plasmonic metal nanoparticles. {\it Nat. Materials} {\bf 14} 567-576 (2015).

\bibitem{Lin_PRB_2008}
Lin, Z., Zhigilei, L. V., Celli, V. Electron-phonon coupling and electron heat capacity of metals under conditions of strong electron-phonon nonequilibrium. {\it Phys. Rev. B} {\bf 77} 075133 (2008).

\bibitem{Ashcroft_1960}
Ashcroft, N. W., Mermin, N. D. Solid State Physics, Saunders College, Philadelphia (1960).

\bibitem{Lide_1993}
Lide, D. R. CRC Handbook of Chemistry and Physics,{\it 74$^{th}$ edition}, Boca Raton (1993).

\bibitem{Rosei_SS_1973}
Rosei, R., Antonangeli, F., Grassano, U. M. d bands position and width in gold from very low temperature thermomodulation measurements. {\it Surf. Sci.} {\bf 37}, 689 (1973).

\bibitem{Christensen_PRB_1971}
Christensen, N. E., Seraphin, B. O., Relativistic Band Calculation and the Optical Properties of Gold. {\it Phys. Rev. B} {\bf 4}, 3321-3344 (1971).

\bibitem{Etchegoin_JCP_2006}
Etchegoin, P. G., Le Ru, E., Meyer, M. An analytic model for the optical properties of gold. {\it J. Chem. Phys.} {\bf 125}, 164705 (2006).

\bibitem{Johnson_PRB_1972}
Johnson, P. B., Christy, R.-W. Optical constants of the noble metals. {\it Phys. Rev. B} {\bf 6} 4370 (1972).

\bibitem{Polli_RSI_2007}
Polli, D., L\"uer, L., Cerullo, G. High-time-resolution pump-probe system with broadband detection for the study of time-domain vibrational dynamics. {\it Rev. Sci. Instrum.} {\bf 78}, 103108 (2007).


\end{thebibliography}

\end{document}


\title{Supplementary Information for: \\
Transient optical symmetry breaking for ultrafast broadband dichroism\\ in plasmonic metasurfaces}

\author{Andrea Schirato$^{1,5^\star}$}

\author{Margherita Maiuri$^{1,2^\star}$}

\author{Andrea Toma$^5$}

\author{Silvio Fugattini$^5$}

\author{Remo Proietti Zaccaria$^{5,6}$}

\author{Paolo Laporta$^{1,2}$}

\author{Peter Nordlander$^{3,4}$}

\author{Giulio Cerullo$^{1,2}$}

\author{Alessandro Alabastri$^{3^\otimes}$}

\author{Giuseppe Della Valle$^{1,2^\otimes}$}

\address{\Polimi,\IFNCNR,\RiceECE,\Rice,\IIT,\China}

\maketitle

\newpage
\section*{Static response}

First of all, polarization resolved linear extinction measurements have been performed to ascertain the isotropic optical response of the fabricated square array of gold nanocrosses. The measured transmission spectrum of the sample in the experimentally available wavelength range ($\sim$530-980~nm) is shown in Fig.~S1a. These data have been exploited to refine the electromagnetic FEM analysis of the structure. 

We assumed an ideal configuration made of an infinite 2D array of the square unit cell reported in Fig.~1c (see main text). The geometrical parameters (cf. main text of the paper) have been finely adjusted within the uncertainty range estimated by SEM images of the sample (Fig.~1a), in order to achieve a best fit with the static measurements of Fig.~S1a. The results of our simulations are shown in Fig.~S1b.

\begin{figure*}[htp!]
\includegraphics[width=7cm]{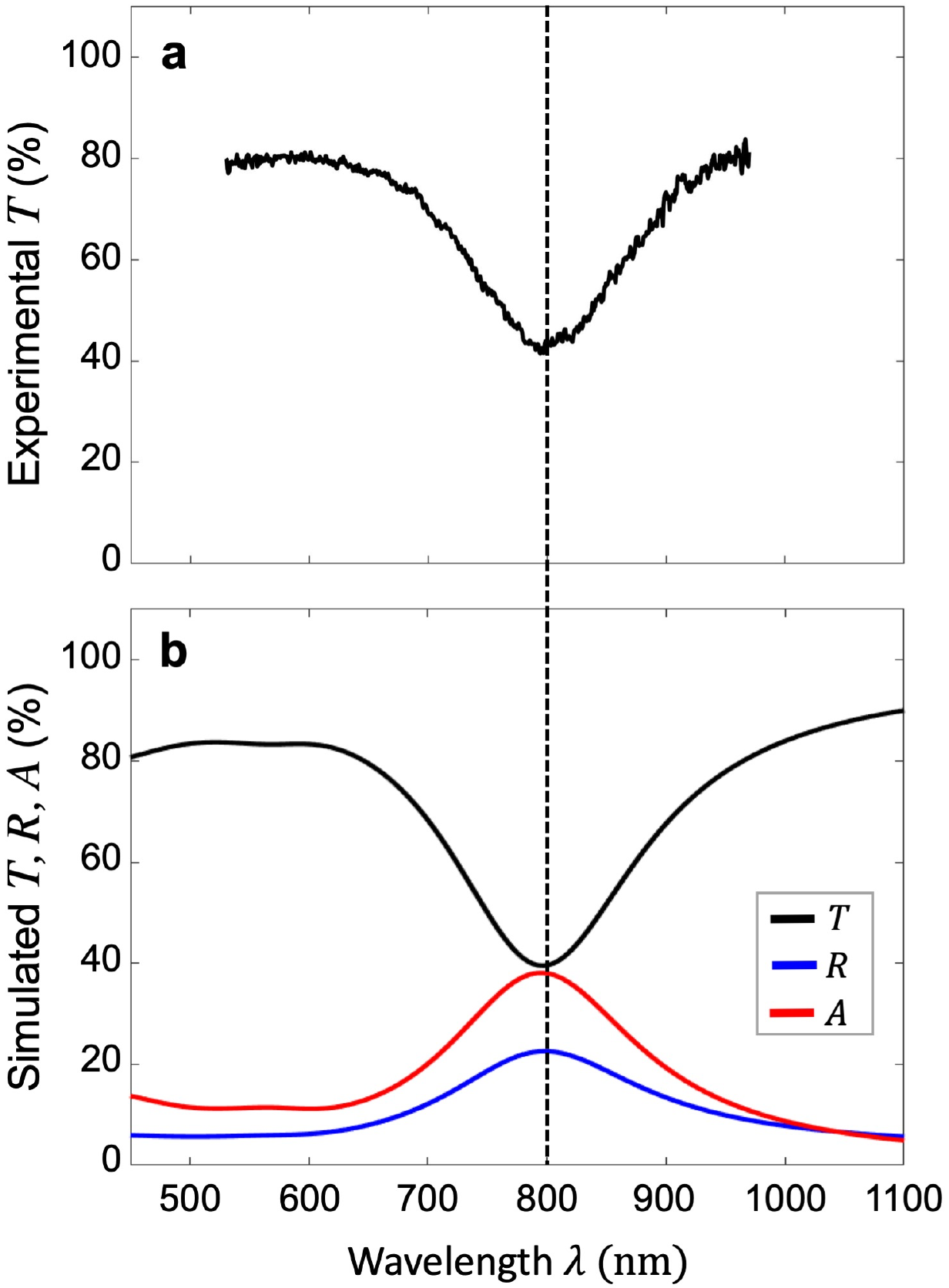}
\caption{{\bf Static optical response of the nanocross metasurface. a}, Measured transmission spectrum of the fabricated sample. {\bf b}, Simulated optical response of the unperturbed Au nanocross array. Note that the plasmonic resonance of the structure dominates its transmission (black), reflection (blue) and absorption (red) spectra.
}
\end{figure*}

Note the good agreement between experimental and simulated transmission (black curves in Fig.~S1a and S1b, respectively), especially in terms of the spectral position and width of the plasmonic resonance dip. In particular, a slight blue-shift of less than $5$ nm is retrieved in the simulations for optimized values of the geometrical parameters. This shift explains the choice of slightly different wavelengths for the cross-sections of Fig.~4c-d.

Note also that a quantitative agreement is also retrieved in terms of the depth of the transmission dip, even though the simulations retrieve a 5\% higher extinction at resonance.

\section*{Space-time dependent optical perturbation}

As mentioned in the main text, the numerical solution of the I3TM, describing the spatio-temporal dynamics of the internal energetic degrees of freedom of the plasmonic nanoparticles (see Eqs.~3-5 and Methods in the main text), is the starting point to compute the corresponding inhomogeneous optical perturbation. Such perturbation in the metallic nanocross, expressed in terms of the spectrally-dispersed complex-valued permittivity, $\varepsilon=\varepsilon(\vec{r}, t, \lambda)$, is retrieved starting from the evolution of $N$, $\Theta_e$ and $\Theta_l$ according to the well-established nonlinearity model briefly discussed in the Methods. Yet, the novelty introduced by the I3TM here reported is represented by the fact that since the rate equations (Eqs.~3-5) are solved locally, the permittivity modulation inherits both the spatial and the temporal dependences from the aforementioned energetic quantities and, as a three-dimensional non-uniform field, it follows a different dynamics in each point of the nanoparticle.
Fig.~S2 highlights such aspects of the I3TM by showing the permittivity spectrum, both in its real and imaginary parts, at different pump-probe time delays and in different points of the structure.

\begin{figure*}[htp!]
\includegraphics[width=16cm]{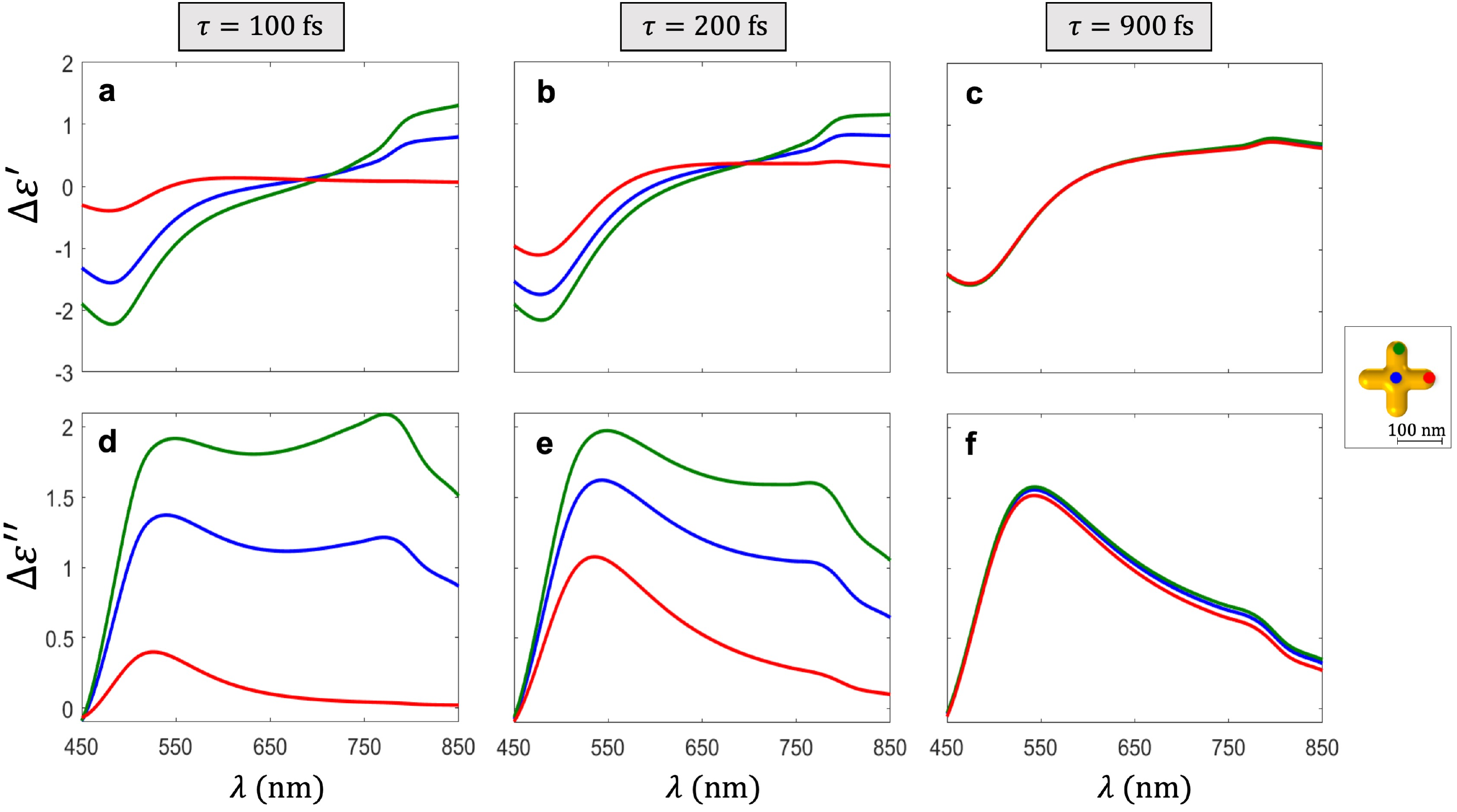}
\caption{{\bf Space-time dependent Au nonlinear permittivity.} Numerically computed pump-induced nonlinear permittivity modulation spectra in different points of the Au nanocross and at different time delays. {\bf a - c}, From left to right: spectra of the real part permittivity variation, $\Delta \varepsilon^\prime$, at $\tau=100$~fs, $\tau=200$~fs, and $\tau=900$~fs, respectively, evaluated at three characteristic positions visualized in the inset of panel (c). {\bf d - f,} Same as a-c but for the imaginary part permittivity variation, $\Delta \varepsilon^{\prime \prime}$.
}
\end{figure*}

\noindent
A time evolution of the permittivity dispersion, non-trivially driven by the interplay of both the nonthermal and thermal hot carriers broadband contributions, is observed in each point of the nanoparticle. Interestingly, Fig.~S2 highlights the fundamental origin of the ultrafast symmetry breaking that we here report. On a sub-picosecond time scale the pump-driven optical perturbation exhibits an inhomogeneous pattern in space. Such non-uniform distribution for $\varepsilon(\vec{r},t,\lambda)$ creates then the ultrafast polarisation-sensitivite response theoretically predicted and experimentally observed. Indeed, the symmetry breaking temporal window, during which the spatial pattern of permittivity remains inhomogeneous, is mainly driven by the nonthermal carriers relaxation and the thermal hot carriers diffusion processes. The characteristic time scales of such energy dynamics, completed within a few hundreds of femtoseconds, entail the purely ultrafast nature of the dichroism we report. By exploiting these dynamics, the effect considered vanishes well before the complete return of the metaatoms to thermodynamic equilibrium. Fig. S2c and S2f, provide a clearcut evidence of this aspect: at a pump-probe delay of $\tau=900$ fs the system is still strongly excited (the curves of $\Delta \varepsilon$ are definitely different from zero). However, at the same time delay homogenisation is achieved and optical symmetry restored (the curves referring to different positions fold and local inhomogeneities collapse), which closes the symmetry breaking window of the dichroic response.

\bibliography{MEP_BIB}